\documentclass[reprint,amsmath,amssymb,pra,twocolumn,longbibliography]{revtex4-2}
\usepackage{graphicx}
\usepackage{color}
\usepackage{bm}
\usepackage{amsmath}
 \usepackage{longtable}

 \def\prb{Phys. Rev. B}
 
 \def\pra{Phys. Rev. A}
 \def\prl{Phys. Rev. Lett.}

\begin{document}

\title{Real transmission and reflection zeros of periodic structures
  \\ with a bound state  in the continuum}

\author{Lijun Yuan}
\email{ljyuan@ctbu.edu.cn}
\author{Mingyang Zhang}
\affiliation{College of Mathematics and Statistics, Chongqing Technology and Business University, Chongqing, 
China \\ Chongqing Key Laboratory of Social Economic and Applied Statistics, Chongqing Technology and Business University, Chongqing, China}

\author{Ya Yan Lu}
\affiliation{Department of Mathematics, City University of Hong Kong, Kowloon, Hong Kong, China} 
\date{\today}

\begin{abstract}
For lossless periodic structures with a proper symmetry, the
transmission and reflection spectra often have peaks and dips that
are truly $100\%$ and $0\%$, respectively. The full peaks
and zero dips typically appear near resonant frequencies, and they 
are robust with respect to structural perturbations that
preserve the required symmetry. However, current theories on the existence
of full peaks and zero dips are incomplete and difficult to use. For
periodic structures with a bound 
state in the continuum (BIC), we present a new theory on the existence
of real transmission and reflection zeros that correspond to the zero
dips in the transmission and reflection spectra. Our theory is
relatively simple, complete, and easy to use. Numerical examples are
presented to validate the new theory. 
\end{abstract}

\maketitle

\section{Introduction}


For periodic structures sandwiched between 
two homogeneous media, the transmission and reflection spectra often
have interesting and useful features. The peaks and dips can be very
sharp. A peak and a dip may appear close to each other and form an
asymmetric line shape. The study on ``anomalous'' transmission and
reflection has a long
history~\cite{wood,fano41}. It is widely accepted that the rapid
change from a peak to a dip, the ``anomaly'' first observed by
Wood~\cite{wood}, is in fact a particular case of Fano
resonance~\cite{fano41,hessel,popov86,fan03}.
The asymmetric line shape is formed from the interference between the resonant  
and non-resonant wave field  components~\cite{fan03}.
The resonant wave field component is caused by the excitation of an
eigenmode of the periodic structure satisfying an outgoing radiation
condition. The eigenmode is either a resonant mode with a complex
frequency or a leaky mode with a complex
propagation constant. The non-resonant field component
exists in a direct passway. In case there is no
direct passway, the spectra exhibit a Lorentzian line shape with only a
single peak or dip~\cite{fan03}. 


For lossless periodic structures, the peaks and dips in the
transmission and reflection spectra can actually be $100\%$
and $0\%$, respectively. Popov {\it et al.}~\cite{popov86} first realized
that structural symmetry is important to the appearance of  full peaks and
zero dips. As functions of frequency and wavenumber, the reflection
and transmission coefficients vanish at their corresponding zeros
(which are complex in general). For lossless periodic structures with
a proper symmetry, the transmission/reflection zeros are either
real or form complex conjugate
pairs~\cite{popov86, nevi95,fehre02,gipp05,blan16,yuan19}. A real transmission zero 
corresponds to a zero dip in the transmission spectrum and a full peak
in the reflection spectrum. Since a simple zero cannot be turned to a
complex conjugate pair by a small perturbation, the zero dips and full
peaks in the transmission/reflection spectra are robust with respect
to small structural perturbations that preserve the required
symmetry~\cite{yuan19}.  However, even for structures with the required
symmetry, the appearance of a real transmission/reflection zero (near a
resonance) is not guaranteed.  To the best of our knowledge,
there is no general theory on the existence of full peaks or zero  dips. 


Shipman and Tu~\cite{shipman12} developed a theory on the existence of
full peaks and zero dips for symmetric lossless periodic structures with a bound state in
the continuum (BIC). A BIC is a guided mode that decays exponentially
in the homogeneous media surrounding the periodic structure, and it
exists in the radiation continuum, namely, there are propagating plane
waves in the homogeneous media having the same frequency and wavenumber as the
BIC~\cite{hsu13,hsu16,kosh19,azzam,sad21}. Importantly, a BIC is a special
point in a band of resonant modes. The theory of Shipman and Tu is
applicable to resonant modes near a BIC. It is mathematically rigorous, but the
technical conditions are specified on quantities that are difficult to
calculate. They identified a generic 
condition under which a real transmission zero and a real reflection
zero likely exist near the frequency of the BIC, 
and they also studied a non-generic case. 


It should be pointed that the temporal coupled-mode theory
(TCMT)~\cite{fan03,wang18,zhao19} can approximate full peaks and zero dips in 
transmission and reflection spectra.
In a recent work~\cite{wu22}, we showed that a direct approximation to
the exact scattering matrix gives the same approximate 
transmission/reflection spectra as the TCMT. 
However, the approximate formulas have limitations
in accuracy, are valid only under proper conditions, and do not
provide a rigorous justification for the existence of real
transmission and reflection zeros. 


In this paper, we present a new  theory on the existence of real
transmission and reflection zeros. Similar to the work of Shipman and
Tu~\cite{shipman12}, our theory is applicable to lossless
periodic structures with a BIC, but the symmetry requirement in our study is less
restrictive. Moreover, our results are relatively simple and more
complete, and the technical conditions are only specified on directly computable
quantities.
The rest of this paper is organized as follows. In
Sec.~II, we recall the definitions and basic properties
for scattering matrices, resonant modes, and BICs.
In Sec.~III, we give a brief summary for the theories of Popov {\it
  et al}~\cite{popov86} and Shipman and Tu~\cite{shipman12}. 
In Sec.~IV, we present our new theory with details on the assumptions,
derivations, and results, and also discuss the difference with
existing theories. To valid our new theory, we present numerical
examples in Sec.~V. The paper is concluded with a
brief discussion in Sec.~VI.

\section{Background}


We consider a lossless two-dimensional (2D) periodic structure that is invariant in
$z$, periodic in $y$ with period $L$, and sandwiched between two
homogeneous media given for $x > D$ and $x < -D$, respectively. The
dielectric function $\varepsilon(x,y)$ of the structure is real and periodic
in $y$, and $\varepsilon(x,y) = \varepsilon_0 \ge 1$ for $|x| > D$. 
In the homogeneous media for $|x| >D$, we specify two  time-harmonic
$E$-polarized plane incident 
waves with a positive angular frequency
$\omega$ and real wave vectors $(\pm \alpha, \beta)$ satisfying
\begin{eqnarray}
  \label{cond_beta}
  &&   -\frac{\pi}{L} < \beta \le \frac{\pi}{L}, \\
  \label{cond_omega}
&&    |\beta| < \frac{\omega}{c} \sqrt{ \varepsilon_0} < \frac{2\pi}{L}
   - |\beta|, \\
&&   \label{def_alpha}
   \alpha = \sqrt{ (\omega/c)^2 \varepsilon_0 - \beta^2}, 
\end{eqnarray}
where $c$ is the speed of light in vacuum and $\alpha > 0$. The $z$ component of the
total electric field, denoted as $u$, can be expanded as
\begin{eqnarray}
\label{leftexpand}
  \nonumber 
&& u(x,y) = b_1^+ e^{ i [ \beta y + \alpha (x + D)]} + b_1^-  e^{ i [ \beta y - \alpha (x + D)]}   \\
&&  \qquad   +  \sum_{j \ne 0} b_{1j} e^{ i \beta_j y + \tau_j (x+D)}, \quad 
   x < -D, \\
\label{rightexpand}  
  \nonumber
  && u(x,y) = b_2^+ e^{ i [ \beta y - \alpha (x - D)]}  + b_2^- e^{ i [ \beta y + \alpha (x - D)]}  \\
  && \qquad +  \sum_{j \ne 0} b_{2j} e^{ i \beta_j y - \tau_j (x-D)},
     \quad  x > D, 
\end{eqnarray}
where $b_1^+$ and $b_2^+$ are the amplitudes of the given incident
waves in the left and right homogeneous media, respectively, $b_1^-$ and
$b_2^-$ are the amplitudes of the outgoing plane waves, 
\begin{equation}
  \label{def_tau}
\beta_j = \beta + 2\pi j/L, \quad 
  \tau_j = \sqrt{ \beta_j^2 - (\omega/c)^2 \varepsilon_0}
\end{equation}
for $j \ne 0$, $\tau_j$ is positive, $b_{1j}$ and $b_{2j}$ are the
amplitudes of the evanescent waves.  
The scattering matrix $S$ satisfies 
\begin{equation}
  \label{Smatrix}
  S
  \begin{bmatrix}
    b_1^+ \cr b_2^+
  \end{bmatrix}
  =
  \begin{bmatrix}
    b_1^- \cr b_2^-
  \end{bmatrix} 
\end{equation}
for any $b_1^+$ and $b_2^+$. We write down the entries of $S$ as
\begin{equation}
  S =
  \begin{bmatrix}
    r & \tilde{t}\  \cr 
    t & \tilde{r}
  \end{bmatrix}, 
\end{equation}
where $r$ and $t$ are the reflection and transmission coefficients
for  the left incident wave, $\tilde{r}$ and $\tilde{t}$ are those of
the right incident wave. 

Clearly, the scattering matrix  $S$ depends on both $\omega$ and
$\beta$. The definition of $S$ can be extended to complex $\omega$ by
analytic continuation. For a real $\beta$, $S$ satisfies
\begin{eqnarray}
\label{unitarity} 
  S^{-1} (\omega, \beta) &=& S^* (\overline{\omega}, \beta), \\
  \label{transpose}
 S^{\sf T} (\omega, \beta) &=& S(\omega, -\beta),
\end{eqnarray}
where $\overline{\omega}$ is the complex conjugate of $\omega$,
$S^{\sf T}$ is the transpose of $S$, and
$S^* (\overline{\omega}, \beta)$ is the conjugate transpose of
$S(\overline{\omega}, \beta)$~\cite{popov86,yuan19}. 
Equations (\ref{unitarity}) and (\ref{transpose}) are related to
energy conservation  and reciprocity, respectively. Notice that if 
$\omega$ is real, $S$ is unitary;  and if $\beta \ne 0$, $S$ is
typically non-symmetric. If the periodic structure has a proper
symmetry, the scattering matrix can be further
simplified~\cite{popov86}. Specifically, we have three cases
\begin{enumerate}
\item[(a).] If $\epsilon(x,y) = \epsilon(-x, -y)$, then $r = \tilde{r}$; 
\item[(b).] If $\epsilon(x,y) = \epsilon(x,-y)$, then $t = \tilde{t}$;
\item[(c).]  If $\epsilon(x,y) = \epsilon(-x, y)$, then $t = \tilde{t}$ and
$r = \tilde{r}$.  
\end{enumerate}

Without incident waves, the periodic structure can support Bloch
eigenmodes given by 
\begin{equation}
  \label{bloch}
  u(x,y) = e^{ i \beta y} \phi(x,y)  
\end{equation}
where $\beta$ is the Bloch wavenumber and $\phi$ is periodic in $y$
with period $L$. For $x  \to \pm \infty$, an eigenmode should either
decay exponentially or radiate out power to infinity (i.e. satisfy the
outgoing radiation condition). For a real $\beta$ satisfying
(\ref{cond_beta}) and the real part of $\omega$ satisfying
(\ref{cond_omega}), expansions (\ref{leftexpand}) and
(\ref{rightexpand}) are applicable to an eigenmode, provided  that we set
$b_1^+=b_2^+=0$. Notice that $b_1^-$ and $b_2^-$ are the coefficients
of outgoing waves. Since the structure is passive and
non-absorbing (i.e. $\varepsilon$ is real and positive), the existence of
outgoing waves is only possible when $\omega$ has a
nonzero imaginary part, so that the eigenmode decays with
time. Such an eigenmode with a real $\beta$ and a complex $\omega$, and
satisfying the outgoing radiation condition, is a resonant mode (also
called a resonant state or quasi-normal mode)~\cite{fan02,amgad}. The resonant modes
form bands. Each band is given by a dispersion relation
$\omega = \omega_\star(\beta)$, where $\omega_\star(\beta)$ is a
complex-valued continuous function of a real variable $\beta$. Notice that $\alpha$
and $\tau_j$ given in (\ref{def_alpha}) and (\ref{def_tau}) are
complex for the resonant modes. Under special
circumstances,  both $b_1^-$ and $b_2^-$ are zero, then $\omega$ is
real,  the eigenmode decays exponentially as $x \to \pm \infty$, and
it is  a guided mode above the light line, also called a bound state
in the continuum (BIC). Note that a BIC is a special point in a band of resonant
modes. If we denote the wavenumber and frequency of a BIC by
$\beta_\dagger$ and $\omega_\dagger$, respectively, then 
$\omega_\dagger = \omega_\star( \beta_\dagger)$. 

Since the inverse of the scattering
matrix $S$ maps $[b_1^-, b_2^-]^{\sf T}$ to $[b_1^+, b_2^+]^{\sf T}$,
and $b_1^+=b_2^+=0$ for a resonant mode, we have
\begin{equation}
  \label{eq:2}
  S^{-1}( \omega_\star, \beta) 
  \begin{bmatrix}
    b_1^- \cr b_2^-
  \end{bmatrix}
  =
  S^* (\overline{\omega}_\star, \beta) 
  \begin{bmatrix}
    b_1^- \cr b_2^-
  \end{bmatrix}
  =
  \begin{bmatrix}
    0 \cr 0
  \end{bmatrix}.
\end{equation}
This above implies that $\omega_\star$ is a pole of $S$, and $\overline{\omega}_\star$
is a zero of $S$. The entries of $S$ can also vanish at special values of $\omega$ and
$\beta$. 
We assume the left transmission coefficient $t= t(\omega, \beta)$ has
a zero set given by a function $\omega=\omega_t^\circ(\beta)$.
Namely, $ t( \omega_t^\circ(\beta), \beta) = 0$ for any $\beta$. Similarly, 
the left reflection coefficient $r$ has a zero set given by a function 
$\omega = \omega_r^\circ (\beta)$. 
For a fixed $\beta$, we call 
$\omega_t^\circ(\beta)$ a transmission zero and
$\omega_r^\circ(\beta)$ a reflection zero. 
In general, for a real
$\beta$, the transmission and reflection zeros are complex. We are
concerned with the existence of real transmission and reflection zeros.

\section{Existing theories}

The possible existence of real transmission and reflection zeros in symmetric
periodic structures was first investigated by Popov {\it et
  al.}~\cite{popov86}. These authors identified the three different symmetries (listed in
Sec.~II) for which $t=\tilde{t}$, $r=\tilde{r}$,  or both
$t=\tilde{t}$ and $r=\tilde{r}$, respectively; and showed that if $t
= \tilde{t}$, then either a transmission
zero is real or there is a pair of complex conjugate
transmission zeros. The case for $r=\tilde{r}$ is similar. This 
implies that a real transmission or reflection zero for a symmetric
periodic structure is robust with respect to structural perturbations
that preserve the symmetry. However, the possibility of complex conjugate
pairs cannot be ruled out, and it remains 
unclear why real transmission and reflection zeros are widely observed
in practice. 

If for a real $\beta$, the periodic structure has a high-$Q$
nondegenerate resonant mode with a complex frequency $\omega_\star =
\omega_0 - i \gamma$ where $\gamma \ll \omega_0$,  and there is no other
resonant modes  near $\omega_\star$, then the reflection and 
transmission coefficients can be written as 
\begin{equation}
  \label{def_ab}
  r(\omega) = \frac{a (\omega)}{\omega - \omega_\star}, \quad 
  t(\omega) = \frac{b(\omega)}{\omega - \omega_\star}, 
\end{equation}
where $a$ and $b$ are analytic functions of $\omega$ near 
$\omega_\star$. 
Since $t$ and $r$ are bounded by $1$ in magnitude, if $\omega$ is
close to $\omega_0$, $|a|/\omega_0$ and $|b|/\omega_0$ must also be
small, but this does not imply that $a$ or $b$ (thus $r$ or $t$) must
have a single zero in a domain that is symmetric about the real axis
and contains $\omega_\star$. 
Consequently,  it is not possible to predict the existence of real
transmission and reflection zeros by considering only  the symmetry of
the structure and the resonant modes.  

Shipman and Tu~\cite{shipman12}
considered periodic structures with a
reflection symmetry in  $x$ [case (c) of Sec.~II], and showed that
the existence of real transmission and reflection zeros is a generic
phenomenon. Their theory is restricted to structures with a BIC
and involves technical conditions are rather difficult to verify. 
More specifically, Shipman and Tu defined an operator $A$ so that the
diffraction problem of Sec.~II becomes
\begin{equation}
  \label{opA}
  A u = p, 
\end{equation}
where both $A$ and $p$ depend on $\omega$ and $\beta$, and $p$
is related to the  incident wave. They further considered the linear eigenvalue 
problem  
\begin{equation}
  \label{lineig}
  A  v = \lambda v, 
\end{equation}
where $\lambda$ is an eigenvalue depending on both $\omega$ and $\beta$,
and $\lambda$ vanishes at the BIC point $(\omega_\dagger, \beta_\dagger)$. 
The technical conditions are 
specified using $\lambda$, $\lambda r$ and $\lambda t$ and their partial 
derivatives at $(\omega_\dagger, \beta_\dagger)$.

The theory of Shipman and Tu is difficult to use in 
practice, because $A$ is abstractly defined without an explicit
expression, and the eigenvalue problem Eq.~(\ref{lineig}) does not
have a clear physical interpretation. Moreover, since Shipman and Tu
only considered case (c), it is not clear  whether their method can be
extended to cases (a) and (b). In our view, a simpler and more
intuitive theory on the existence of real  transmission and reflection
zeros is highly desirable.

\section{New theory}

In this section, we present a new theory to clarify the conditions for
the existence of real transmission and reflection zeros. We consider a lossless
periodic structure as described in Sec.~II, and 
assume there is a BIC with frequency $\omega_\dagger$ and Bloch
wavenumber $\beta_\dagger$ satisfying conditions (\ref{cond_beta}) and
(\ref{cond_omega}). The BIC is supposed to be a special point in a
band of nondegenerate resonant modes with dispersion relation $\omega =
\omega_\star(\beta) = \omega_0 (\beta) - i \gamma(\beta)$, where
$\omega_0$ and $-\gamma$ are the real and imaginary parts
of $\omega_\star$, and they satisfy
$\gamma(\beta_\dagger) = 0$,  $\omega_0(\beta_\dagger) =
\omega_\star( \beta_\dagger) = \omega_\dagger$, and
$\gamma(\beta) > 0$ for $\beta$ near but not equal to $\beta_\dagger$.

Our theory depends on the analyticity of a few functions. First of
all, $\omega_\star$ is an analytic function of real variable $\beta$~\cite{shipman12}. Therefore,
$\omega - \omega_\star(\beta)$, as a function of complex variable  $\omega$
and real variable $\beta$, is analytic in $\omega$ and $\beta$. It is known
that the reflection and transmission coefficients, $r$ and $t$, as
functions of $\omega$ and $\beta$, are not continuous at
$(\omega_\dagger, \beta_\dagger)$~\cite{shipman12}. However, we claim
that the two functions 
$a$ and $b$ given by 
\begin{eqnarray*}
  \label{defnewa}
  a(\omega, \beta) &=& [ \omega - \omega_\star(\beta) ] r(\omega, 
                       \beta), \\
  \label{defnewb}  
  b(\omega, \beta) &=& [ \omega - \omega_\star(\beta) ] t(\omega, 
  \beta), 
\end{eqnarray*}
are analytic in $\omega$ and $\beta$ in a neighborhood of the BIC point
$(\omega_\dagger, \beta_\dagger)$. This implies that the singularity
of $r$ and $t$ at $(\omega_\dagger, \beta_\dagger)$ can be removed by
multiplying $\omega - \omega_\star(\beta)$.  Notice that 
$a(\omega_\dagger, \beta_\dagger) = b(\omega_\dagger, \beta_\dagger) =
0$. The above proposition is different from our statement about
Eq.~(\ref{def_ab}), where $\beta$ is fixed, the dependence on $\beta$
is suppressed, $\mbox{Im}(\omega_\star) \ne 0$, 
and $a$ and $b$ are analytic in a single complex variable $\omega$.
The analyticity of $a$ and $b$ can be established following the proof
for the analyticity of $\lambda t$ and $\lambda r$ by Shipman and
Tu~\cite{shipman12}, where $\lambda$ is the eigenvalue defined in
Eq.~(\ref{lineig}).

Similarly, for the right reflection and transmission coefficients, the
two functions 
$\tilde{a} = [\omega - \omega_\star(\beta)] \tilde{r}$ and
$\tilde{b} = [\omega - \omega_\star(\beta)] \tilde{t}$ are analytic in
$\omega$ and $\beta$ near $(\omega_\dagger, \beta_\dagger)$. The
scattering matrix can be written as
\begin{equation}
  \label{scaledS}
  S(\omega, \beta)
  = \frac{1}{\omega - \omega_\star(\beta)}
  \begin{bmatrix}
    a(\omega, \beta) & \tilde{b}(\omega, \beta) \cr b(\omega, \beta) & \tilde{a}(\omega, \beta)
  \end{bmatrix}
\end{equation}
for $\omega \ne \omega_\star(\beta)$, where $a$, $b$, $\tilde{a}$ and
$\tilde{b}$ are analytic in $\omega$ and $\beta$, and vanish at
$(\omega_\dagger, \beta_\dagger)$. 

Next, we consider the function 
\begin{equation}
  \label{def_f}
  f(\omega, \beta)     = \frac{ \omega - \omega_\star(\beta)}{\omega -
    \overline{\omega}_\star(\beta)} \, \det S(\omega, \beta).
\end{equation}
For a fixed $\beta \ne \beta_\dagger$, $\omega_\star = \omega_\star(\beta)$ is
complex with a nonzero imaginary part. Since the resonant mode is
nondegenerate, $\omega_\star$ is a simple pole of $\det S$ and
$\overline{\omega}_\star$ is a simple zero of $\det S$. Therefore, for
the fixed $\beta$, $f$ given in Eq.~(\ref{def_f}) is an analytic
function of $\omega$. For $\beta = \beta_\dagger$ and $\omega \ne
\omega_\dagger$, we have $f(\omega, \beta) = \det S (\omega,
\beta_\dagger)$. It is known that for a fixed $\beta$, the reflection
and transmission coefficients are continuous in $\omega$. Therefore,
we define $f(\omega_\dagger, \beta_\dagger)$ by 
\begin{equation}
  \label{def_fBIC}
  f(\omega_\dagger, \beta_\dagger) =
\lim_{\omega \to \omega_\dagger} \det S(\omega, \beta_\dagger) =
  \det S(\omega_\dagger, \beta_\dagger).
\end{equation}
Numerical results suggest that
$f(\omega, \beta)$ is an analytic
function of $\omega$ and $\beta$ for $\omega$ 
near $\omega_\dagger$ and $\beta$ near $\beta_\dagger$. Unfortunately,
a formal mathematical proof is currently not available.

Using Eq.~(\ref{unitarity}), we can easily show that
\begin{equation}
  \label{fone}
  \overline{f}(\overline{\omega}, \beta)
  f(\omega, \beta) = 1, 
\end{equation}
where $\overline{f}(\overline{\omega}, \beta)$ is the complex
conjugate of $f(\overline{\omega}, \beta)$. In particular, 
if $\omega$ is real, then $|f(\omega, \beta)|=1$. 
Since $|f(\omega_\dagger, \beta_\dagger)|=1$,
we can choose a small neighborhood (denoted as $\Omega$) of $(\omega_\dagger,
\beta_\dagger)$,  and choose a proper branch cut
to define a complex square root function, so that 
$g(\omega, \beta) = \sqrt{f(\omega, \beta)}$ is nonzero and analytic
(in $\omega$ and $\beta$) on 
$\Omega$. With this function $g$, we can rewrite the scattering
matrix as
\begin{equation}
  \label{def_RT}
  S(\omega, \beta) = \frac{g(\omega,\beta)}{\omega -
    \omega_\star(\beta)}
  \begin{bmatrix}
    R(\omega, \beta) & \tilde{T}(\omega, \beta) \cr
    T(\omega, \beta) & \tilde{R}(\omega, \beta)
  \end{bmatrix}, 
\end{equation}
where $(R, T, \tilde{R}, \tilde{T}) =(a, b, \tilde{a}, \tilde{b})/g$.
Clearly, $R$, $T$,  $\tilde{R}$ and $\tilde{T}$ 
are analytic functions of $\omega$ and $\beta$ on $\Omega$, and
they vanish at $(\omega_\dagger, \beta_\dagger)$. These functions are
introduced,  because they have useful properties. Using
Eq.~(\ref{unitarity}), we can show that  
\begin{eqnarray}
\label{complex_T}
  \tilde{T}(\omega,\beta) &=& - \overline{T}(\overline{\omega},\beta), \\
  \label{complex_R}
\tilde{R}(\omega,\beta) &=& \overline{R}(\overline{\omega},\beta).
\end{eqnarray}
%
%
In addition, it is also easy to verify that
\begin{equation}
  \label{pomega}
  \frac{\partial R}{\partial \omega} (\omega_\dagger, \beta_\dagger)
  = \frac{r_{\dagger}}{g_{\dagger}},  
  \quad
  \frac{\partial T}{\partial \omega} (\omega_\dagger, \beta_\dagger) 
  = \frac{t_{\dagger}}{g_{\dagger}},  
\end{equation}
where $r_\dagger  = r(\omega_{\dagger}, \beta_\dagger)$,
$t_\dagger  = t(\omega_\dagger, \beta_\dagger)$ and 
$g_\dagger = g(\omega_{\dagger}, \beta_\dagger)$. 

For case (a) of Sec.~II, we have $r=\tilde{r}$, and thus 
\begin{equation}
  \label{Rcaseb}
  R(\omega,\beta) = \overline{R}(\overline{\omega},\beta). 
\end{equation}
This implies that $R(\omega, \beta)$ is a real analytic function of
two real variables $\omega$ and $\beta$ near $
(\omega_\dagger, \beta_\dagger)$. Since $R (\omega_\dagger,
\beta_\dagger)=0$, we can analyze the zero set of $R$ using the
partial derivatives of $R$ at $(\omega_\dagger, \beta_\dagger)$.
\begin{enumerate}
\item $\partial_\omega R \ne 0$ at $(\omega_\dagger, \beta_\dagger)$.  According to
  Eq.~(\ref{pomega}), this condition is equivalent to $r_\dagger \ne
  0$.  In the $\omega$-$R$ plane,  $R = R(\omega,
  \beta_\dagger)$ is a curve passing through zero at $\omega_\dagger$
  with a nonzero slope. When $\beta$ is slightly changed, the curve is
  slightly shifted. The new curve,  given by $R = R(\omega, \beta)$, 
  still passes through zero near $\omega_\dagger$. More precisely, according to
  the implicit function theorem, when $\partial_\omega R$ is nonzero
  at $(\omega_\dagger, \beta_\dagger)$, $R(\omega, \beta) = 0$ can be
  uniquely solved near $\beta_\dagger$, and the solution is a
  function $\omega = \omega_r^\circ (\beta)$, such that
$\omega_r^\circ(\beta_\dagger) = \omega_\dagger$ and $R(
\omega_r^\circ (\beta), \beta) = 0$ for $\beta$ near
$\beta_\dagger$. Therefore, if $r_\dagger \ne 0$ and $\beta$ is close
to $\beta_\dagger$, there is a real reflection zero
$\omega_r^\circ(\beta)$ near $\omega_\dagger$. This is the generic
case and $r_\dagger \ne 0$ is the generic condition.

\item At $(\omega_\dagger, \beta_\dagger)$, $\partial_\omega R = 0$, 
  $\partial^2_\omega R \ne 0$, $\partial^2_\beta R \ne 0$, 
  $\partial^2_\omega R $ and $\partial^2_\beta R $ have the same 
  sign.  This is a non-generic case.  The condition $\partial_\omega R =0$ implies 
  $r_\dagger = 0$,  and thus $\partial_\beta R=0$ at $(\omega_\dagger, 
  \beta_\dagger)$. This is the case for which $(\omega_\dagger, \beta_\dagger)$ is a local 
  extremum of $R$.  The function $R(\omega, \beta)$ is nonzero for 
  real $(\omega, 
  \beta)$ near but not equal to $(\omega_\dagger, 
  \beta_\dagger)$. There is no real reflection zero for $\beta$ near
  but not equal to $\beta_\dagger$. 

\item At $(\omega_\dagger, \beta_\dagger)$, $\partial_\omega R = 0$, 
  $\partial^2_\omega R \ne 0$, $\partial^2_\beta R \ne 0$, 
  $\partial^2_\omega R $ and $\partial^2_\beta R $ have the opposite
  sign. This is another non-generic case.  
  In the $\omega$-$R$ plane, $R = R(\omega, \beta_\dagger)$
  obtains a local minimum (or maximum) at $\omega_\dagger$ and is
  exactly zero at $\omega_\dagger$. For $\beta$ near $\beta_\dagger$,
  the minimum (or maximum) of the curve $R = R(\omega, \beta)$ is
  negative (or positive), and the curve passes zero at two values of
  $\omega$ near $\omega_\dagger$. On the surface given by $R =
  R(\omega, \beta)$, $(\omega_\dagger,  \beta_\dagger)$ is a saddle
  point. For $\beta$ near but not equal to $\beta_\dagger$, there are
  real two reflection zeros near $\omega_\dagger$. 
\end{enumerate}
There are other non-generic cases. For example, when 
$\partial_\omega R = 0$, $\partial_\omega^2 R \ne 0$, we can consider the
case $\partial_\beta^2 R = 0$ and $\partial_\beta^3 R \ne 0$. This
leads to one real reflection zero for $ \beta < \beta_\dagger$ (or
$\beta > \beta_\dagger$) and
no real reflection zero   for $\beta > \beta_\dagger$ (or 
$\beta < \beta_\dagger$).  We can also
consider new cases assuming $\partial_\omega R = \partial^2_\omega R = 
0$ and $\partial^3_\omega R \ne 0$. However, these additional
non-generic cases are too special and difficult to find in practice. 

For case (b), we have $t=\tilde{t}$, thus 
\begin{equation}
  \label{Tcasea}
T(\omega,\beta) = - \overline{T}(\overline{\omega},\beta).
\end{equation}
Clearly, if $\omega$ is real, $T$ is pure imaginary. Following the
approach for analyzing $R$, we also identify 
the generic case and two main non-generic cases as follows.
\begin{enumerate}
\item $\partial_\omega T \ne 0$ at $(\omega_\dagger, \beta_\dagger)$,
  equivalent to $t_\dagger \ne 0$. This is the generic case and the
  generic condition is $t_\dagger \ne 0$.   For
  $\beta$ near $\beta_\dagger$, there is one transmission zero
  $\omega_t^\circ (\beta)$ near $\omega_\dagger$.
\item At $(\omega_\dagger, \beta_\dagger)$, $\partial_\omega T = 0$, 
  $\partial_\omega^2 T \ne 0$, 
  $\partial_\beta^2 T \ne 0$,   and 
  $( \partial_\omega^2 T) (\partial_\beta^2 T) < 0$. This is a non-generic 
  case. $\mbox{Im}(T)$ has a local extremum at $(\omega_\dagger, 
  \beta_\dagger)$. There is no real transmission zero for $\beta$ near 
  $\beta_\dagger$ and $\beta \ne \beta_\dagger$.
\item At $(\omega_\dagger, \beta_\dagger)$, $\partial_\omega T = 0$, 
  $\partial_\omega^2 T \ne 0$, 
  $\partial_\beta^2 T \ne 0$,   and 
  $( \partial_\omega^2 T) (\partial_\beta^2 T) >  0$. This is another  non-generic 
  case.
    On the surface given by $\mbox{Im}(T)$ as a function of real
  variables $\omega$ and $\beta$, $(\omega_\dagger,    \beta_\dagger)$
  is a saddle point.   For any  $\beta$ near 
  $\beta_\dagger$ and $\beta \ne \beta_\dagger$, there are two real
  transmission zeros.   
\end{enumerate}

For case (c), we have both $t=\tilde{t}$ and $r=\tilde{r}$. Therefore,
the above results on transmission and reflection zeros for cases (a) and 
(b), respectively,  are valid for case (c).

\section{Numerical examples}

\begin{figure}[http]
  \centering 
   \includegraphics[scale=0.7]{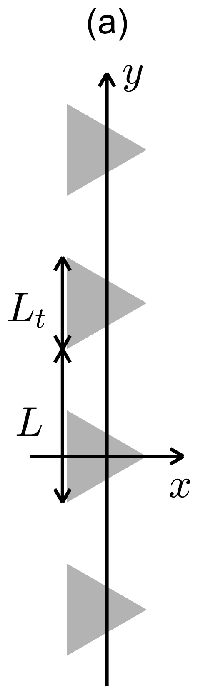}
   \includegraphics[scale=0.7]{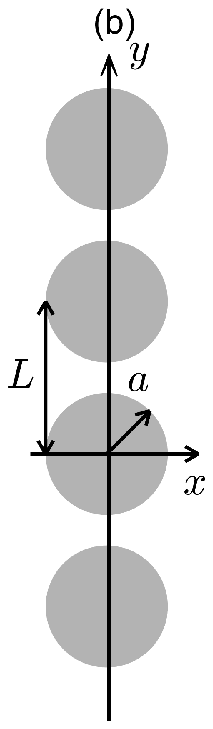}
 \caption{Two periodic arrays of cylinders with period $L$ in the $y$
   direction: (a) triangular cylinders with side length $L_t$; (b)
   circular cylinders with radius $a$.} 
 \label{figforstru}
\end{figure}

To validate our theory, we present several numerical
examples involving periodic arrays of dielectric cylinders. 
In Fig.~\ref{figforstru},
we depict  two periodic arrays with triangular and circular cylinders,
respectively. The arrays are periodic in $y$ with period
$L$. The triangular cylinders have one surface parallel to the $y$ axis, and
their cross sections are equilateral triangles with side length
$L_t$. The radius of the circular cylinders is $a$. All cylinders have
the same dielectric constant $\varepsilon_1=10$ and are surrounded by
air (with dielectric constant $\varepsilon_0=1$).


Our theory is for periodic structures with a BIC. It is
well known that many different BICs may exist in a periodic array of
dielectric cylinders~\cite{shipman03,port05,mari08,bulg14,yuanJPB,bulg17pra,hu18,yuan20_2,amgad21}.   Our first example is for the array of  
triangular cylinders shown in Fig.~\ref{figforstru}(a).
For $L_t = 0.45L$, we found a BIC with frequency
$ \omega_\dagger  = 0.6875 (2\pi c/L)$ and wavenumber 
$\beta_\dagger = 0$. Since $\beta_\dagger$ is zero and the electric field
is an odd function of $y$, this BIC is an anti-symmetric standing
wave.  The transmission coefficient at 
$(\omega_\dagger, \beta_\dagger)$ satisfies 
$|t_\dagger| = 0.7530$.
Since the periodic array has a 
reflection symmetry in $y$ and  $t_\dagger \neq  0$, there should be a 
real transmission zero $\omega_t^\circ$ for $\beta$ near $\beta_\dagger = 0$.
\begin{figure}[http]
  \centering 
  \includegraphics[scale=0.75]{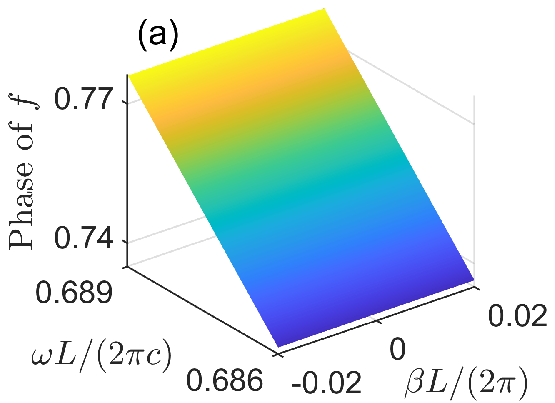} 
 \includegraphics[scale=0.75]{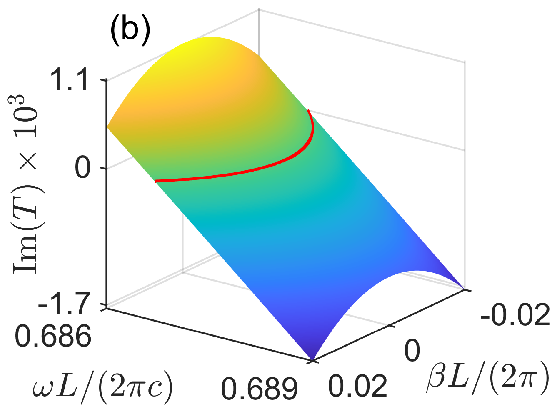}   
 \includegraphics[scale=0.75]{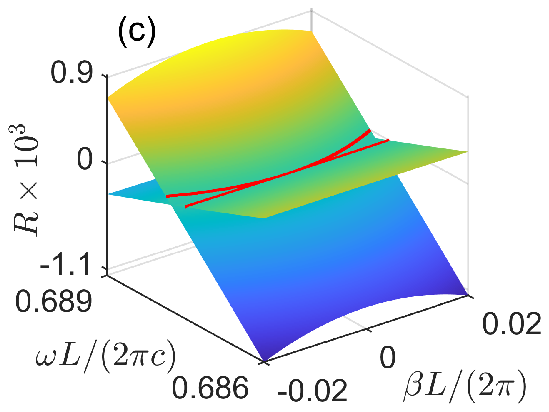}
  \includegraphics[scale=0.79]{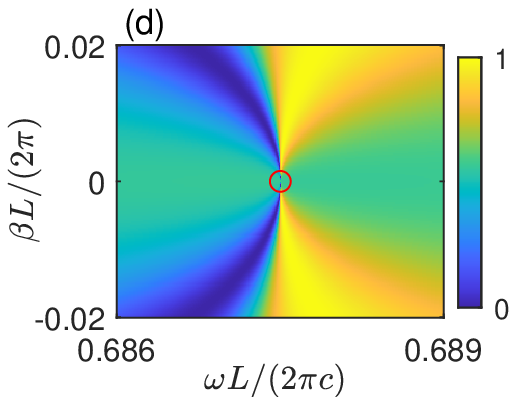}
 \includegraphics[scale=0.8]{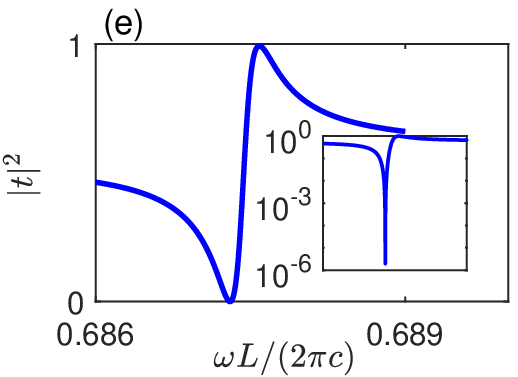}
 \includegraphics[scale=0.8]{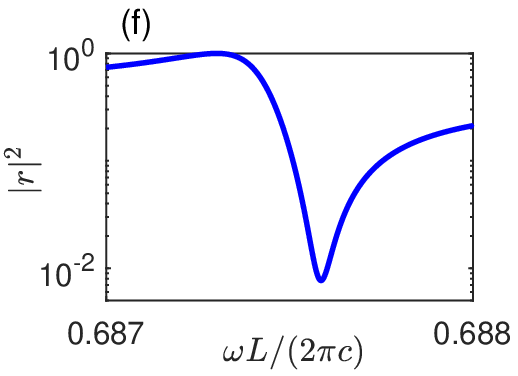}
  \caption{Transmission and reflection near a BIC in a periodic array
    of triangular cylinders with $L_t=0.45L$: 
    (a) the phase of $f$ in unit of $\pi$,  (b) $\mbox{Im}(T)$, (c)  $\mbox{Re}(R)$ and
    $\mbox{Im}(R)$,  (d) $|t|^2$, (e)  $|t|^2$ for $\beta =
    0.01 (2\pi/L)$ (with a logarithmic plot in the inset).  (f): 
     $|r|^2$ for $\beta = 0.01 (2\pi/L)$ (in a logarithmic 
    scale).  The solid red curves in (b) and (c)  correspond to $T= 0$,
    $\mbox{Re}{(R)} = 0$ and $\mbox{Im}{(R)} = 0$, respectively. The
    circle in (d) corresponds the BIC.} 
  \label{fig_ASW1}
\end{figure}
In Fig.~\ref{fig_ASW1}(a)-(d), we show 
the phase $f$, $T$, $R$ and
$|t|^2$ for $(\omega, \beta)$ near $(\omega_\dagger, \beta_\dagger)$. 
Our theory relies on the
analyticity  of function $f$ in $\omega$ and $\beta$. Since $|f(\omega,\beta)|=1$ for
real $(\omega, \beta)$, we
show the phase of $f$ (in unit of $\pi$) in Fig.~\ref{fig_ASW1}(a).
The periodic array has a symmetry corresponding to case (b) of
Sec.~II, thus $T$ is pure imaginary and $R$ is complex for real $(\omega, \beta)$. In
Fig.~\ref{fig_ASW1}(b) and (c), we show $\mbox{Im}(T)$, $\mbox{Re}(R)$
and $\mbox{Im}(R)$, and highlight their zero sets by the solid red
curves. Notice that the two curves for $R$ touch tangentially at
$(\omega_\dagger, \beta_\dagger)$. In Fig.~\ref{fig_ASW1}(d), we show
transmittance $|t|^2$ as a 
function of $\omega$ and $\beta$, where the BIC point $(\omega_\dagger,
\beta_\dagger)$ is marked by a small circle. To show peaks and dips more
clearly, we  plot transmission and reflection spectra ($|t|^2$ and
$|r|^2$ as functions of $\omega$) for $\beta = 0.01 (2\pi/L)$ in
Fig.~\ref{fig_ASW1}(e) and (f), respectively.
There is indeed a zero dip (corresponding to the real transmission
zero)  in the transmission spectrum, but the
dip in the reflection spectrum is not zero, and the peak in the
transmission spectrum is not $100\%$.

The second example is for a periodic array of circular cylinders with
$a=0.3L$.
\begin{figure}[t]
  \centering 
  \includegraphics[scale=0.75]{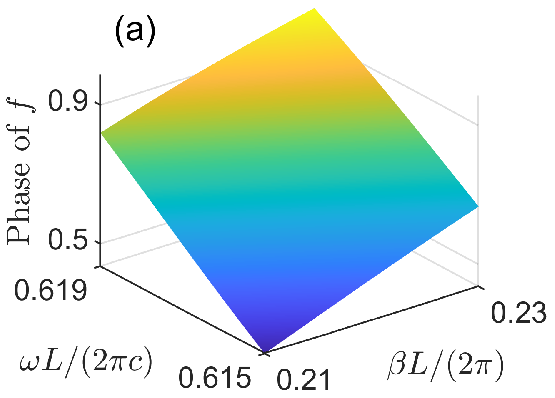}  
  \includegraphics[scale=0.75]{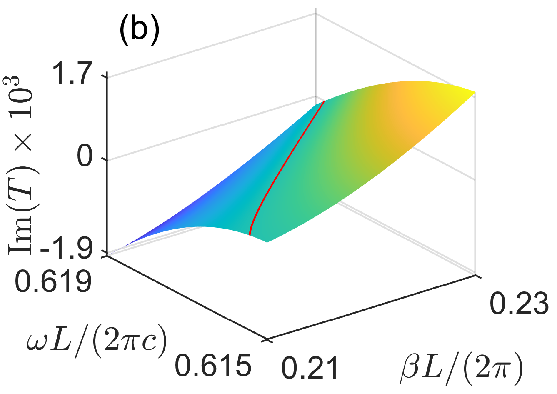} 
  \includegraphics[scale=0.75]{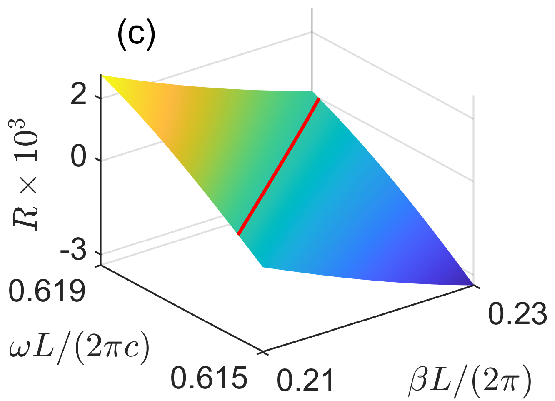}
  \includegraphics[scale=0.8]{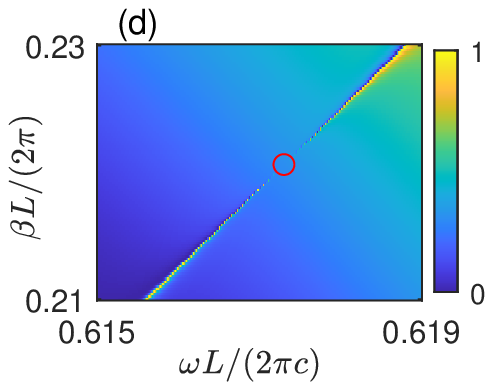}
  \includegraphics[scale=0.8]{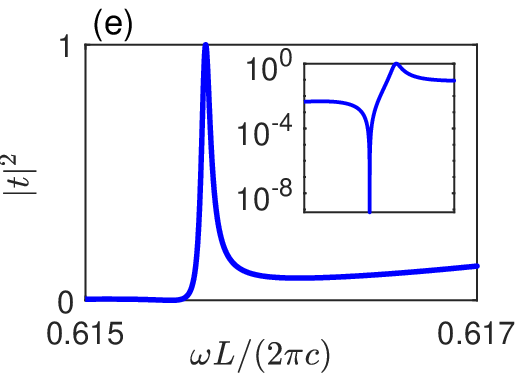}
  \includegraphics[scale=0.8]{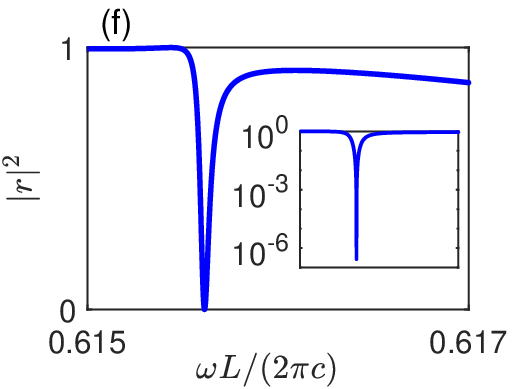}
  \caption{Transmission and reflection near a BIC in a periodic array
    of circular cylinders with radius $a=0.3L$:  (a)  the phase of $f$ in unit of $\pi$,   
    (b) $\mbox{Im}(T)$, (c) $R$,    (d) $|t|^2$ with the BIC marked by ``$\circ$''.
    (e)  $|t|^2$ for     $\beta  =     0.21 (2\pi/L)$,
    (f) $|r|^2$ for $\beta  =     0.21 (2\pi/L)$.
    The solid red curves in (b) and (c) correspond to $T=0$ and $R=0$,
    respectively. Insets in
    (e) and (f) are logarithmic plots. }
  \label{fig_PBIC1}
\end{figure}
In this array, there is a propagating BIC with 
$\beta_\dagger = 0.2206 (2\pi/L)$  and $\omega_\dagger = 0.6173 (2\pi
c/L)$. The reflection coefficient at 
the BIC point satisfies $ |t_\dagger| = 0.5568$. 
The symmetry of the structure corresponds to case (c) of
Sec.~II. Since  $t_\dagger \neq 0$ and $r_\dagger \neq 0$, both
transmission and reflection coefficients have one real  
zero for $\beta$ close to $\beta_*$.
In Fig.~\ref{fig_PBIC1}(a)-(d),
we show the phase of $f$, $\mbox{Im}(T)$, $R$ and $|t|^2$ as functions of real $\omega$ and $\beta$.  
The real transmission and reflection zeros 
form curves in the $\omega$-$\beta$ plane and shown as
the solid red curves in Fig.~\ref{fig_PBIC1}(b) and (c),
respectively. The transmittance $|t|^2$ is shown in
Fig.~\ref{fig_PBIC1}(d) for $(\omega, \beta)$ near $(\omega_\dagger,
\beta_\dagger)$. To show the peaks and dips more clearly, we plot
transmission and reflection spectra for $\beta = 0.21 (2\pi/L)$ in
Fig.~\ref{fig_PBIC1}(e) and (f),  respectively. The zero dips in the
spectra correspond to the real transmission and reflection zeros. 

\begin{figure}[t!]
  \centering 
  \includegraphics[scale=0.75]{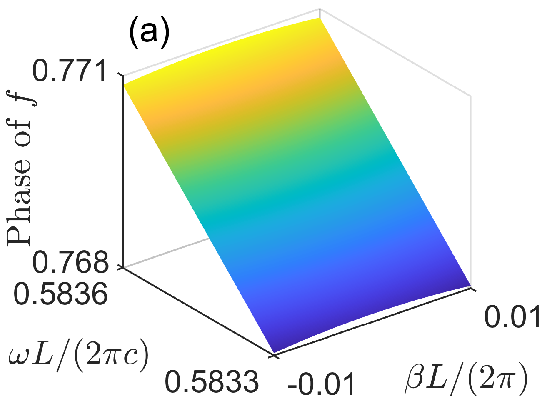}  
    \includegraphics[scale=0.75]{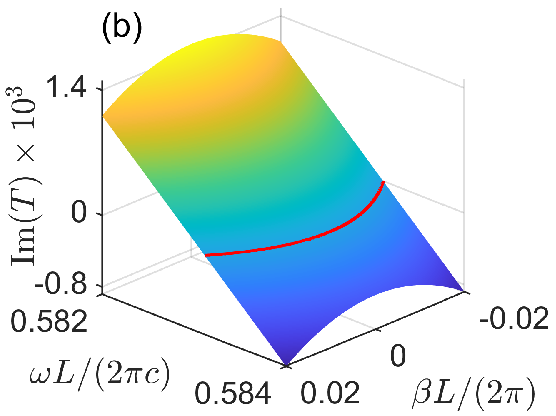} 
  \includegraphics[scale=0.75]{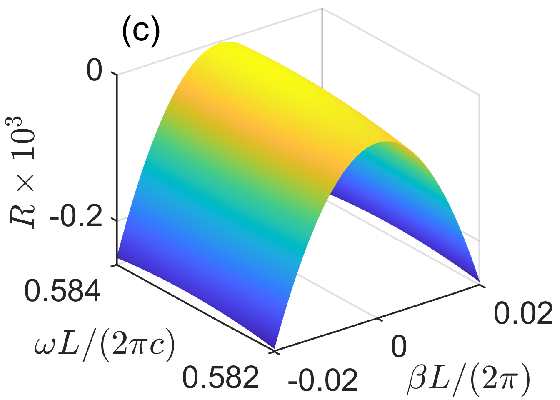}
  \includegraphics[scale=0.8]{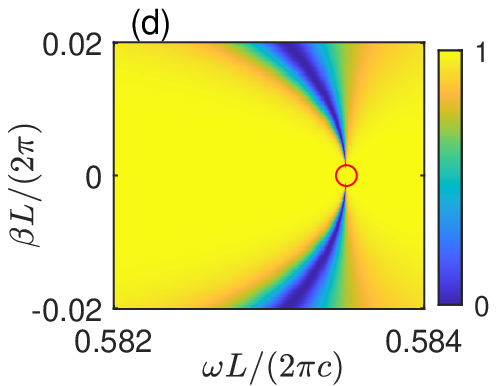}
  \includegraphics[scale=0.8]{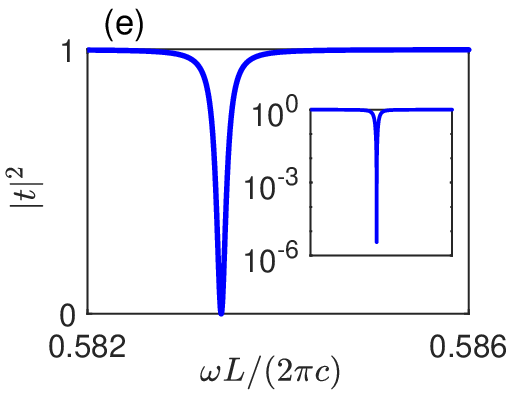}
  \includegraphics[scale=0.8]{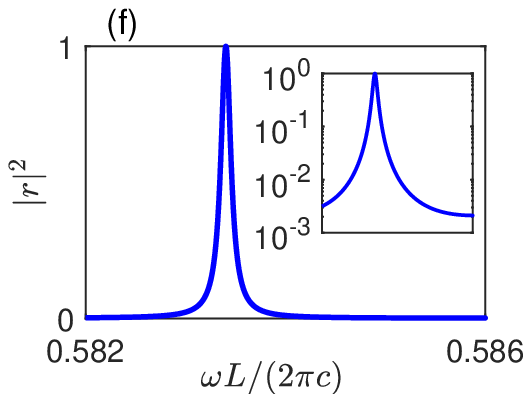}
  \caption{Transmission and reflection near a BIC in a periodic array
    of circular cylinders with radius $a=0.2074L$: (a) the phase of
    $f$ in unit of $\pi$, (b) $\mbox{Im}(T)$, (c) $R$, (d) $|t|^2$ with the BIC marked
    by ``$\circ$'', (e) transmission spectrum and (f) reflection
    spectrum for $\beta =  0.01 (2\pi/L)$.
    The solid red curve in (b) corresponds to $T=0$. The sets in 
    (e) and (f) are logarithmic plots.}
  \label{fig_ASW3}
\end{figure}

The first two examples cover only generic cases with 
$t_\dagger \ne 0$ and $r_\dagger \ne 0$. The third example is designed
to illustrate a non-generic case. The structure is still a periodic array of circular
cylinders. Many BICs in the array exist continuously with respect to
radius $a$. For $a = 0.2074L$, there is a BIC (an anti-symmetric
standing wave) with $\beta_\dagger = 0$, $\omega_\dagger = 0.5835
(2\pi c/L)$, and $r_\dagger = 0$.
In Fig.~\ref{fig_ASW3}, 
we show the phase of  $f$, imaginary part of $T$, $R$ and $|t|^2$ as
functions of $\omega$ and $\beta$, and transmission and reflection
spectra for $\beta = 0.01 (2\pi/L)$. Since $|t_\dagger | = 1$, there
is a real transmission zero for each $\beta$ near $\beta_\dagger =
0$, and it corresponds to the red curve in Fig.~\ref{fig_ASW3}(b). 
On the other hand, 
$(\omega_\dagger, \beta_\dagger)$ is a local maximum point of $R$ as
shown  in Fig.~\ref{fig_ASW3}(c), and
there are no real reflection zeros for $\beta$ near $\beta_\dagger$. 
For $\beta = 0.01 (2\pi/L)$, the transmission spectrum, shown in
Fig.~\ref{fig_ASW3}(e), has an inverted Lorentzian line shape and a
zero dip. 
The reflection spectrum, shown in Fig.~\ref{fig_ASW3}(f), has a full
peak and no dip in the frequency range. 

\begin{figure}[t]
  \centering
  \includegraphics[scale=0.75]{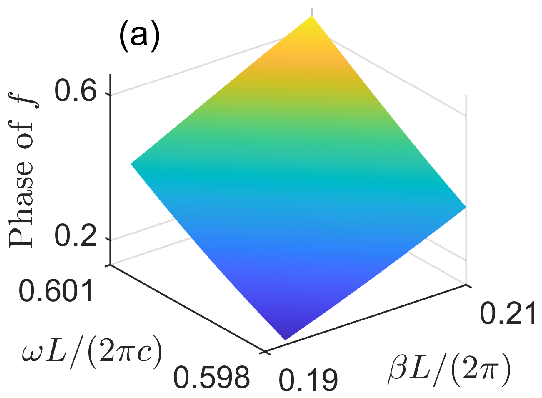} 
  \includegraphics[scale=0.75]{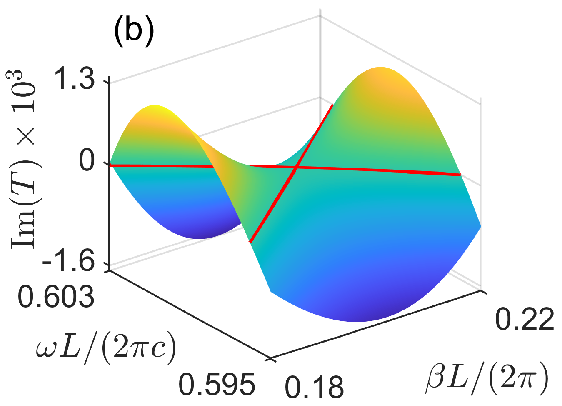} 
  \includegraphics[scale=0.75]{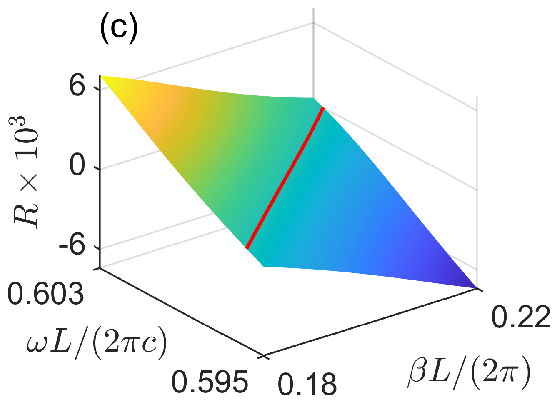}
  \includegraphics[scale=0.8]{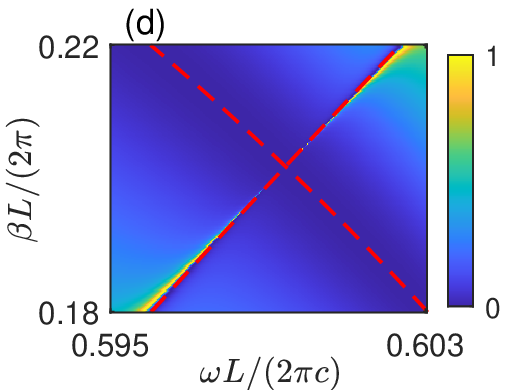}
  \includegraphics[scale=0.8]{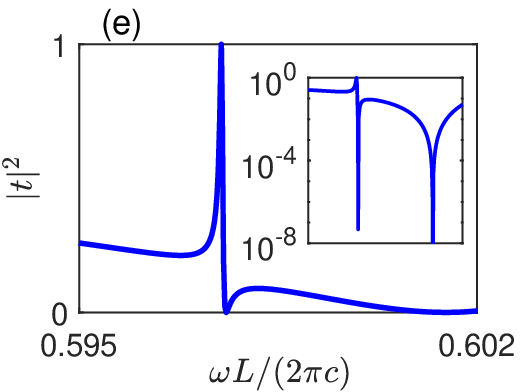}
  \includegraphics[scale=0.8]{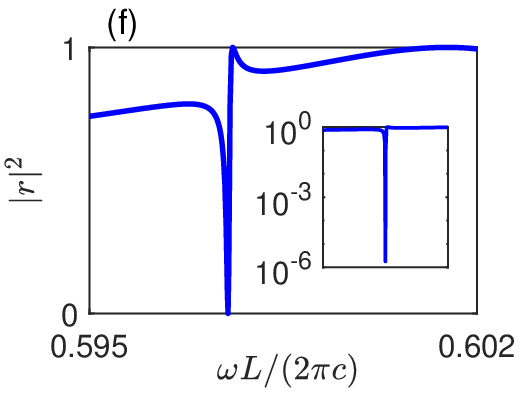}
  \caption{Transmission and reflection near a BIC in a periodic array
    of circular cylinders with radius $a = 0.3087L$: 
    (a) the phase of $f$ in unit of $\pi$, (b)  $\mbox{Im}(T)$, (c) $R$, (d) $|t|^2$,
    (e) $|t|^2$ for $\beta  = 0.19 (2\pi/L)$,
    (f) $|r|^2$ for $\beta  = 0.19 (2\pi/L)$.
    The solid red curves in (b) and (c), and the dashed red curves in
    (d), are the real zero sets of $T$ (or $t$) and $R$.
    The insets in (e) and (f) are logarithmic plots.}
  \label{fig_PBIC2}
\end{figure}

The fourth example is also designed to exhibit a non-generic case. It
is well known that a periodic array of circular cylinders can support
propagating BICs (with a nonzero Bloch wavenumber) that depend on
the radius $a$ continuously.  For radius $a = 0.3087L$, we found a 
propagating BIC with
$\beta_\dagger = 0.2018 (2\pi/L)$ and $\omega_\dagger = 0.5994 (2\pi
c/L)$, and $t_\dagger = 0$. In Fig.~\ref{fig_PBIC2}(a)-(d),
we show the phase of $f$, $\mbox{Im}(T)$, $R$, and $|t|^2$, respectively, as
functions of real $(\omega, \beta)$. It is clear that
 $(\omega_\dagger, \beta_\dagger)$ is saddle point of $\mbox{Im}(T)$. 
For each $\beta$ near $\beta_\dagger$, the transmission coefficient has two real 
 zeros. As shown in Fig.~\ref{fig_PBIC2}(b), the zero set of $T$ form two intersecting 
 curves in the $\omega$-$\beta$ plane. The real transmission zeros are
 also shown in Fig.~\ref{fig_PBIC2}(d) as the red dashed curves. 
In Fig.~\ref{fig_PBIC2}(e) and (f), we show the 
 transmission and reflection spectra,  respectively, for $\beta = 0.19
 (2\pi/L)$. As shown clearly in the logarithmic plot (in the inset), there are
 two zero dips in the transmission spectrum. Since $|r_\dagger| =1$,
 there is one real reflection zero for each $\beta$ near
 $\beta_\dagger$. The zero set of $R$ is shown as the red curve in 
Fig.~\ref{fig_PBIC2}(c). The reflection spectrum of
Fig.~\ref{fig_PBIC2}(f) shows clearly a single zero dip.

The two examples above illustrate the non-generic cases where 
$(\omega_\dagger, \beta_\dagger)$ is an extremum point of $R$ and a
saddle point of $\mbox{Im}(T)$, respectively. We have also found
numerical examples (still for a periodic array of circular cylinders)
where $(\omega_\dagger, \beta_\dagger)$ is an extremum point of
$\mbox{Im}(T)$ or a saddle point of $R$. Since the results are quite
similar, we skip these examples.

\section{Conclusion}

For lossless periodic structures with a proper symmetry, the
transmission and reflection spectra often have peaks and dips that are 
truly $100\%$ and zero, respectively. When the transmission and 
reflection coefficients are considered as functions of the frequency
$\omega$, they vanish at their corresponding zeros, but the zeros 
are complex in general. A zero dip in the transmission/reflection spectrum
corresponds to a real zero of the transmission/reflection
coefficient. Existing theories on real transmission/reflection zeros 
have limitations and may be difficult to use~\cite{popov86,shipman12}. 
In this paper,  a relatively simple theory on the existence
(and nonexistence) of real transmission/reflection zeros is developed.
The key step is to scale the transmission and 
reflection coefficients, $t$ and $r$,  to $T$ and $R$, such that for
structures with a proper symmetry, $T$ is a pure imaginary analytic function
and $R$ is a real analytic function of real $\omega$ and $\beta$. We identified 
the generic case and two non-generic cases, for which the number of
real transmission/reflection zeros  is 1, 0, or 2, respectively.
The non-generic cases appear when the reflection or transmission
coefficient at $(\omega_\dagger, \beta_\dagger)$ (the BIC frequency
and wavenumber) vanishes, and when $(\omega_\dagger, \beta_\dagger)$
is either an extremum point or a saddle point of $R$ or
$\mbox{Im}(T)$. 
Our theory is validated by numerical examples involving periodic arrays of
dielectric cylinders. 

It should be pointed out that our theory relies on the analyticity of function $f$
given in Eq.~(\ref{def_f}), but a rigorous mathematical proof is not 
available. In addition, the theory is only applicable to periodic structures
with a BIC. The transmission and reflection spectra are those for
incident waves with a wavenumber $\beta$ near $\beta_\dagger$.
For resonant scattering problems without wavenumber $\beta$, for
example, the scattering problem for a local defect in a closed
waveguide, the transmission and reflection spectra can be approximated
using the transmission and reflection coefficients at the real resonant
frequency~\cite{wu22}. However, it is not clear whether this approximate
theory can be used to establish the existence of real
transmission/reflection zeros rigorously.

\section*{Acknowledgments}
 The authors acknowledge support from the Natural Science Foundation
 of Chongqing, China (Grant No. cstc2019jcyj-msxmX0717),  the Graduate Student Research Innovation Project of Chongqing, and the
 Research Grants Council of Hong Kong Special Administrative Region,
 China (Grant No. CityU 11304619).

\end{document}